\documentclass[aps,prd,twocolumn,groupedaddress]{revtex4}
\usepackage{graphicx}% Include figure files
\usepackage{dcolumn}% Align table columns on decimal point
\usepackage{bm}% bold math
\usepackage{amsmath}
\usepackage{amssymb}
\usepackage{color}
\usepackage{slashed}

\begin{document}

\title{Submarine paradox softened}

\author{Hrvoje Nikoli\'c}\email{hnikolic@irb.hr}
\affiliation{Theoretical Physics Division, Rudjer Bo\v{s}kovi\'{c} Institute, P.O.B. 180, HR-10002 Zagreb, Croatia.}

\begin{abstract}
In Supplee's submarine paradox, a naive argument based on Lorentz contraction
leads to a contradiction that a fast submarine should sink in the water's reference frame
but float in the submarine's reference frame.
Due to the submarine's rigidity constraints, it is not easy to resolve the paradox in a manifestly covariant form. 
To simplify the problem, we consider a version of the paradox in which 
one fluid moves through another fluid. 
An analysis of ideal relativistic fluids in a weak gravitational field 
shows that the moving fluid has a larger pressure and hence sinks, 
in agreement with known results for the rigid submarine.
\end{abstract}

%\vspace*{0.5cm}
%Keywords: 

\maketitle

\section{Introduction} 
 
Consider a submarine at rest with respect to water, having the same density as water so that it neither sinks nor floats.
What happens if the submarine attains a relativistic velocity $v$ in the horizontal direction? 
In the water's rest frame, the submarine should become Lorentz contracted, so it should become 
denser than water and therefore sink. But analogously, in the reference frame where the submarine is at rest, 
it is the water that should become Lorentz contracted and denser, so the submarine should float. But it is a contradiction;
whether the submarine will sink or float cannot depend on the observer. This puzzle, known as the 
submarine paradox, was first studied by Supplee \cite{supplee} and then further analyzed from various points of view in 
\cite{matsas,jonsson,vieira}. They all agreed that the submarine will, in fact, sink.
Physically, this is because buoyancy depends on Earth's gravity, so the 
relevant velocity to consider is the velocity of the submarine  
{\em relative to Earth} (which is independent of the observer's velocity).

One difficulty which is encountered when analyzing the submarine paradox is the fact that the usual Lorentz contraction formula 
$L'=L/\gamma$ is not a transformation law for a tensor. In this formula, $L$ is the proper length, namely
the length of object as seen by observer comoving with the object, $L'$ is the length seen by observer
moving with velocity $v$ relative to the comoving observer, and
\begin{equation}\label{gamma}
\gamma=\frac{1}{\sqrt{1-v^2}} 
\end{equation}
is the usual special-relativistic factor, in units where $c=1$. On the other hand, since the problem involves gravity,
its proper treatment requires the use of general relativity naturally formulated in terms of tensors. 
An argument based on the Lorentz contraction formula
lacks manifest general covariance and looks rather awkward 
from the general-relativistic point of view. To our knowledge, the only 
manifestly covariant treatment of the submarine paradox is the one by Matsas \cite{matsas}. However, the treatment
by Matsas is quite involved, mainly because one needs to implement a constraint that the submarine 
is a rigid object. 
(In fact, even within a special-relativistic context, 
the rigidity constraint may be complicated when one studies 
an object with a non-constant velocity \cite{nik_rod}.)

The motivation for this work is to find a resolution of the submarine paradox 
which, like \cite{matsas}, is relativistically covariant, but at the same time involves simpler 
formalism than \cite{matsas}. Our main idea is to drop the rigidity constraint.
Indeed, if the shape of an object is changed such that its density and volume remain the same
in its own rest frame,
the effect of buoyancy should be the same. Hence the rigidity constraint 
(the constraint that the object's shape and volume should 
remain constant in its own reference frame) seems inessential.
Thus, instead of studying a rigid submarine, we study a fluid, 
first idealized as incompressible with
a fixed constant density $\rho$, and later, more realistically, as a slightly compressible fluid like water.  
This significantly simplifies the problem 
because a manifestly covariant treatment is much simpler for
fluids than for rigid objects. We study a drop of one fluid (called ``ink'') of proper density $\rho$, 
moving horizontally through another fluid (called the ``lake'') of the same proper density $\rho$. 
Here the names ``ink'' and ``lake'' are taken for the sake of intuitive visualization.
(To avoid diffusion of one fluid into another, in a realistic experiment one might want to deal with fluids 
at a temperature close to absolute zero, so water-based fluids would not be the best choice.
Alternatively, perhaps it is possible to have two different immiscible fluids with almost equal densities.) 
As with the submarine, the question is whether the 
moving drop will sink or float. By an elementary yet
manifestly covariant analysis, we find that it will sink.

The paper is organized as follows. In Sec.~\ref{SECpres} we start from the energy-momentum tensor for an ideal
fluid in a weak 
quasi-Newtonian
gravitational field, and derive a formula that determines the enhancement 
of the fluid's pressure due its horizontal velocity $v$. In Sec.~\ref{SECcomov} we 
repeat the same calculation in a comoving frame with respect to which 
the fluid does {\em not} move, and explain why the pressure still gets enhanced due to $v$,
showing that the pressure enhancement is observer independent. 
In Sec.~\ref{SECdrop} we apply those results to a moving incompressible drop to show that the 
pressure enhancement of the drop implies that the drop sinks.
In Sec.~\ref{SECcompr} we discuss the effect of compressibility of a more realistic fluid
and show that it does not significantly affect the results.
In Sec.~\ref{SECfull} we analyze how the pressure enhancement is modified in the  
full weak field limit, beyond the quasi-Newtonian approximation.
Finally, a qualitative discussion of our results is given in Sec.~\ref{SECdisc}.

\section{Pressure of ideal fluid in a weak 
quasi-Newtonian 
gravitational field}
\label{SECpres}

We start from the well known formula for the
energy-momentum tensor of an ideal relativistic fluid  \cite{weinberg}
\begin{equation}\label{enmom}
 T^{\mu\nu}=pg^{\mu\nu}+(p+\rho)u^{\mu}u^{\nu} ,
\end{equation}
where $\rho$ is density of the fluid, $p$ is pressure of the fluid, $u^{\mu}$ is the 4-velocity of the fluid,
and $g^{\mu\nu}$ is the spacetime metric tensor with signature $(-+++)$. 
We work in units $c=1$, so $p$ and $\rho$ have the same units. 
Here $p$ and $\rho$ transform as scalars under general coordinate transformations; 
they are the pressure and energy density as seen by observer comoving with the fluid.
The 4-velocity vector is defined as
\begin{equation}
 u^{\mu}=\frac{dx^{\mu}}{d\tau} ,
\end{equation}
where $\tau$ is the proper time defined by $d\tau^2=-g_{\mu\nu}dx^{\mu}dx^{\nu}$, which implies
\begin{equation}\label{norm}
 g_{\mu\nu}u^{\mu}u^{\nu}=-1.
\end{equation}

To avoid misunderstanding of the notion of relativistic pressure, 
some additional conceptual clarifications may be useful. In non-relativistic physics ``pressure'' 
is usually defined as the force per unit area when the force 3-vector and area 3-vector are parallel. 
However, in relativity such a quantity 
depends on the frame of reference and hence is not a scalar.
In relativistic physics the pressure of a fluid is hence redefined as above
{\em in the local frame of reference in which the fluid is at rest}, which is why the relativistic 
pressure $p$ is a scalar.  
Similarly, the density $\rho$ of the fluid is redefined 
as mass per volume {\em in the local frame of reference in which the fluid is at rest}.
(This is similar to the fact that mass $m$ 
of a relativistic particle is defined as energy in the frame of reference in which the particle is at rest,
which is why $m$ is a scalar.) Indeed, the scalar nature of $p$ and $\rho$ is clear from the tensor nature 
of (\ref{enmom}).

If no external force (except gravity) acts on the fluid, 
then its energy-momentum tensor obeys a covariant conservation law \cite{weinberg}
\begin{equation}
 \nabla_{\nu}T^{\mu\nu}=0 .
\end{equation}
In explicit form it can be written as \cite{weinberg}
\begin{equation}\label{eqp}
 g^{\mu\nu}\partial_{\nu}p+\frac{1}{\sqrt{g}} \partial_{\nu}[\sqrt{g}(p+\rho)u^{\mu}u^{\nu}]
+(p+\rho)\Gamma^{\mu}_{\nu\lambda}u^{\nu}u^{\lambda}=0, 
\end{equation}
where $g$ is the absolute value of the determinant of $g_{\mu\nu}$ and 
\begin{equation}\label{christ}
 \Gamma^{\sigma}_{\lambda\mu}=\frac{g^{\nu\sigma}}{2} 
(\partial_{\lambda}g_{\mu\nu}+\partial_{\mu}g_{\lambda\nu}-\partial_{\nu}g_{\mu\lambda})
\end{equation}
are the Christoffel symbols.

We are interested in the case where the metric $g_{\mu\nu}$ describes a weak gravitational field
near the Earth's surface. 
The full weak field limit will be explored in Sec.~\ref{SECfull}, but here, for simplicity,
we shall work in the weak quasi-Newtonian limit
\begin{eqnarray}\label{metric}
 & g_{00}=-(1+2\phi), &
\nonumber \\
 & g_{0i}=0, \;\;\; g_{ij}=\delta_{ij}, & 
\end{eqnarray}
where $i,j=1,2,3$ are the spatial indices,
\begin{equation}\label{potential}
 \phi(x^{\mu})=\phi_0+g_{\rm acc}x^3 
\end{equation}
is the Newtonian gravitational potential, $x^3=z$ is the vertical coordinate, 
$g_{\rm acc}=9.81\,{\rm m}/{\rm s}^2$ is the gravitational acceleration on Earth,
and $\phi_0$ is an arbitrary constant, which determines the altitude 
at which the potential is taken to be zero. 
In the quasi-Newtonian limit above, only $g_{00}$ is nontrivial, while in the full weak-field limit
(explored in Sec.~\ref{SECfull}) 
$g_{ij}$ are also nontrivial \cite{footnote}. 
What we call the quasi-Newtonian limit is usually called 
the Newtonian limit in the literature, but we add the ``quasi'' prefix to emphasize
an important difference between non-relativistic Newtonian gravity and gravity described by (\ref{metric}).
The metric (\ref{metric}) transforms as a tensor under general coordinate transformations, 
while in non-relativistic Newtonian gravity $\phi$ transforms as a scalar. In fact, the metric (\ref{metric})
is given in a frame of reference in which Earth is at rest. We shall see in Sec.~\ref{SECcomov}  
that in another frame of reference the metric tensor takes a different form, while in non-relativistic Newtonian gravity
that would not be the case.   

There are several reasons for working with the quasi-Newtonian limit instead of the full weak field limit.
First, this better isolates the kinematic essence of our resolution of the submarine paradox,
which involves certain physical quantities 
transforming as scalars, vectors or tensors. Second, other papers on the submarine paradox also work 
in this limit, either explicitly \cite{vieira} or implicitly \cite{supplee,jonsson}. Third, 
the effect of gravity in the submarine paradox can be simulated 
by a uniformly accelerating frame of reference in flat spacetime \cite{matsas} and only $g_{00}$ 
is nontrivial in the local coordinate system of an accelerated observer \cite{mtw}.    

Now suppose that the fluid moves with a horizontal velocity
\begin{equation}\label{v}
\frac{dx^1}{dx^0}=v 
\end{equation}
in the $x^1$-direction. Thus we have
\begin{equation}
 u^1=\frac{dx^1}{d\tau}=\frac{dx^1}{dx^0} \frac{dx^0}{d\tau} =vu^0 ,
\end{equation}
so $u^{\mu}=(u^0,u^1,u^2,u^3)=(u^0,vu^0,0,0)$. From this, normalization (\ref{norm}) 
and the metric (\ref{metric}), we find 
\begin{equation}\label{ug}
 u^{\mu}=\tilde{\gamma}(1,v,0,0) ,
\end{equation}
where
\begin{equation}\label{tgamma}
 \tilde{\gamma}=\frac{1}{\sqrt{|g_{00}|-v^2}}
\end{equation}
is a generalization of (\ref{gamma}).

Alternatively, instead of using $v$ defined by (\ref{v}), one could define $v_{\rm loc}=dx^1/dt_{\rm loc}$, where 
$dt_{\rm loc}=\sqrt{|g_{00}(z)|}dx^0$ is the time measured by a local static observer staying at an arbitrary 
altitude $z$. Hence $v_{\rm loc}$ can be interpreted as the velocity of a material point at $z$,
as seen by an observer at the {\em same} $z$. (For a similar, more general construction see also \cite{bilic}.)  
In contrast, (\ref{v}) can be interpreted as the velocity of a material point at $z$,
as seen by a static observer at the {\em fixed} altitude $z_*$ for which $|g_{00}(z_*)|=1$,
which implies $z_*=-\phi_0/g_{\rm acc}$ due to (\ref{potential}).
The two velocities are related as $v_{\rm loc}=v/\sqrt{|g_{00}(z)|}$, so (\ref{tgamma}) can be written as
\begin{equation}\label{tgamma2}
 \tilde{\gamma}=\frac{1}{\sqrt{|g_{00}|}\sqrt{1-v_{\rm loc}^2}} .
\end{equation}
We emphasize that (\ref{tgamma}) and (\ref{tgamma2}) are obtained in the weak quasi-Newtonian limit (\ref{metric}).
Either $v$ or $v_{\rm loc}$ can be used to parameterize the 4-velocity $u^{\mu}$; choosing one or the other is a matter 
of convenience. For our purposes $v$ is more convenient for the following reason. 
If a rigid body extended in the vertical $z$-direction
moves horizontally, then intuitively one expects that the upper end of the body has the same ``velocity'' 
as its lower end, i.e. that the ``velocity'' does not depend on $z$. The ``velocity'' 
that is constant in that sense is $v$, not $v_{\rm loc}$. 

Having specified the four-velocity and the metric,
Eq.~(\ref{eqp}) can be thought of as an equation that determines $p(x)$. Since the metric (\ref{metric})-(\ref{potential})
and the velocity (\ref{ug})-(\ref{tgamma}) depend only on $x^3=z$, while $\rho$ is either a constant 
(for an incompressible fluid) or satisfies an equation of state $\rho=\rho(p)$,
it follows that $p$ depends only on $x^3=z$. Hence the middle term in (\ref{eqp}), proportional
to an expression of the form $\partial_{\nu}[\cdots u^{\nu}]$, vanishes; 
terms with $\nu\neq 3$ do not contribute
due to the derivative $\partial_{\nu}$, while the term with $\nu=3$ does not contribute because $u^3=0$.
A similar analysis of other terms reveals that (\ref{eqp}) reduces to
\begin{equation}\label{eqp2}
 \partial_{3}p+
(p+\rho)\left[ \Gamma^{3}_{00}u^{0}u^{0}+(\Gamma^{3}_{01}+\Gamma^{3}_{10})u^{0}u^{1}+\Gamma^{3}_{11}u^{1}u^{1} \right]=0 . 
\end{equation}
The relevant Christoffel symbols are computed from (\ref{christ}) for the metric (\ref{metric}) depending only on $x^3$
\begin{eqnarray}\label{Christ}
 & \Gamma^{3}_{00}=-\frac{1}{2}\partial_3g_{00}, &
\nonumber \\
 & \Gamma^{3}_{01}=\Gamma^{3}_{10}=0, \;\;\; \Gamma^{3}_{11}=0 . &
\end{eqnarray}
Hence (\ref{eqp2}) simplifies to
\begin{equation}\label{eqp3}
 \partial_{3}p=(p+\rho) \frac{\partial_3g_{00}}{2} (u^{0})^2 .
\end{equation}
Next, (\ref{metric}) and (\ref{potential}) give
\begin{equation}\label{derg00}
 \frac{\partial_3g_{00}}{2}=-g_{\rm acc} ,
\end{equation}
while (\ref{ug})-(\ref{tgamma}) gives
\begin{equation}
 (u^{0})^2=\tilde{\gamma}^2 ,
\end{equation}
so (\ref{eqp3}) finally becomes
\begin{equation}\label{eqp4}
 \partial_{3}p=-(p+\rho)g_{\rm acc} \tilde{\gamma}^2 .
\end{equation} 

The meaning of (\ref{eqp4}) is most easily understood by considering the nonrelativistic limit
$\rho\gg p$, $\tilde{\gamma}\approx 1$, in which case (\ref{eqp4}) becomes
\begin{equation}\label{nonrel}
 \partial_z p \approx -\rho g_{\rm acc} .
\end{equation}
If the fluid is incompressible so that $\rho$ is constant,
this gives $p(z)\approx p_0-\rho g_{\rm acc} z$, which is  
the familiar nonrelativistic linear dependence of pressure on depth.
This reveals that (\ref{eqp4}) is just a relativistic generalization of the 
nonrelativistic formula (\ref{nonrel}). 
For our purposes, the most important feature of (\ref{eqp4}) is the fact that 
{\em the pressure gradient gets enhanced by the relativistic factor $\tilde{\gamma}^2$ 
that increases with velocity $v$}. 

\section{Pressure analyzed from the comoving frame}
\label{SECcomov}

The result (\ref{eqp4}) has been obtained by doing all calculations in Earth's frame, in which 
the fluid moves with horizontal velocity $v$. 
%The dependence on $v$ is present in the factor
%$\tilde{\gamma}^2$ defined by (\ref{tgamma}). 
Now we want to find the pressure $p$ by doing analogous calculations
in the frame comoving with the fluid. Naively one might argue that in the comoving frame the fluid does
not move, so there can be no $\tilde{\gamma}^2$ in that frame. But that cannot be right.
The pressure $p$ is a scalar, so its value cannot change due to a change of coordinates.
%It must have the factor $\tilde{\gamma}^2$ even in the comoving frame. 
To see where the $\tilde{\gamma}^2$ comes from in the comoving frame,
in this section we perform an explicit calculation.

To define a comoving frame, consider a Lorentz transformation
\begin{eqnarray}\label{lorentz}
 & x'^0=\gamma (x^0-vx^1), \;\;\; x'^1=\gamma (x^1-vx^0), &
\nonumber \\
 & x'^2=x^2, \;\;\; x'^3=x^3 . &
\end{eqnarray}
The inverse transformation is of course
\begin{eqnarray}\label{inverse}
 & x^0=\gamma (x'^0+vx'^1), \;\;\; x^1=\gamma (x'^1+vx'^0), &
\nonumber \\
 & x^2=x'^2, \;\;\; x^3=x'^3 . &
\end{eqnarray}
The 4-velocity in the primed coordinates is then
\begin{equation}\label{ug'}
 u'^{\mu}=\frac{\partial x'^{\mu}}{\partial x^{\alpha}} u^{\alpha}
= \frac{\tilde{\gamma}}{\gamma} (1,0,0,0) ,
\end{equation}
where (\ref{lorentz}) and (\ref{ug}) have been used. We see that all spatial components of $u'^{\mu}$ vanish, 
which shows that $x'^{\mu}$ are indeed comoving coordinates. Similarly, the metric in primed
coordinates is
\begin{equation}
 g'_{\mu\nu}= \frac{\partial x^{\alpha}}{\partial x'^{\mu}} \frac{\partial x^{\beta}}{\partial x'^{\nu}} 
g_{\alpha\beta} ,
\end{equation}
so (\ref{inverse}) and (\ref{metric}) give
\begin{eqnarray}\label{metric'}
 & g'_{00}=-\displaystyle\frac{|g_{00}|-v^2}{1-v^2}=-\frac{\gamma^2}{\tilde{\gamma}^2} , &
\nonumber \\
& g'_{11}= \displaystyle\frac{1-|g_{00}|v^2}{1-v^2} , \;\;\; g'_{22}=g'_{33}=1 , &
\nonumber \\
& g'_{01}= \displaystyle\frac{(1-|g_{00}|)v}{1-v^2} , &
\end{eqnarray}
with other components vanishing.
We see that $g'_{00}\neq g_{00}$, $g'_{11}\neq g_{11}$ and  $g'_{01}\neq g_{01}$, 
i.e. the Lorentz transformation is not an isometry
of the metric in the presence of the static gravitational field described by $|g_{00}|\neq 1$.
(It is possible to redefine the time coordinate to make $g_{00}=-1$ everywhere, but then 
some of the metric components become time dependent so the gravitational field does not longer look static.) 
In particular,
even though we are in a comoving frame, the metric depends on $v$. It is only in the 
absence of gravity, when $|g_{00}|= 1$, that the dependence 
on $v$ in (\ref{metric'}) disappears and the Lorentz transformation becomes an isometry,
namely a transformation for which $g'_{\mu\nu}=g_{\mu\nu}$.

Now we can see what happens with the pressure in the comoving frame. By analysis completely analogous to the one 
that gave (\ref{eqp3}), in the comoving coordinates we obtain
\begin{equation}\label{eqp3c}
 \partial'_{3}p=(p+\rho) \frac{\partial'_3g'_{00}}{2} (u'^{0})^2 ,
\end{equation}
which indeed has the same form as (\ref{eqp3}). Furthermore, $\partial'_{3}=\partial_{3}$ due to (\ref{lorentz}).
The difference between (\ref{eqp3c}) and (\ref{eqp3}) is that $g'_{00}\neq g_{00}$ and
$(u'^{0})^2\neq (u^{0})^2$. However, a comparison of (\ref{ug'}) and (\ref{ug}) reveals that
\begin{equation}\label{comovu}
 (u'^{0})^2=\frac{1}{\gamma^2}(u^{0})^2 ,
\end{equation}
while a similar comparison of (\ref{metric'}) and (\ref{metric})-(\ref{potential}) reveals that
\begin{equation}\label{comovgamma}
 \partial'_3g'_{00}=\gamma^2 \partial_3g_{00} .
\end{equation}
Hence $\partial'_3g'_{00} (u'^{0})^2 = \partial_3g_{00} (u^{0})^2$, so (\ref{eqp3c}) is really the same equation
as (\ref{eqp3}). 
Taking for granted that the boundary condition, such as $p(z_0)=0$, cannot depend on the observer,
it follows that
%
%In other words, 
the pressure $p$ analyzed from the comoving frame is the same as pressure $p$ analyzed from 
the frame in which fluid moves, which indeed must be the case because $p$ is a scalar.

A reader with only special-relativistic intuition may still be confused why the pressure of fluid gets enhanced 
due to $v$ even in the comoving frame in which the fluid does not move. 
%Such a reader may understand intuitively 
%the $1/\gamma^2$ factor in (\ref{comovu}) but may fail to understand intuitively the $\gamma^2$ %factor in (\ref{comovgamma}).
In special relativity, the Minkowski metric tensor does not change under Lorentz transformations, so intuitively one may wrongly 
expect that $g'_{00}=g_{00}$. Indeed, if there was no $\gamma^2$ factor 
in (\ref{comovgamma}) to cancel the $1/\gamma^2$ factor in (\ref{comovu}), there would be no 
pressure enhancement due to $v$ in the comoving frame. This demonstrates that special-relativistic 
intuition based on an invariant metric tensor is not sufficient to properly understand the submarine paradox.
Instead, one must develop general relativistic intuition 
and note that the metric tensor is also a quantity that depends on the reference frame. 

\section{Soft submarine}
\label{SECdrop}

Now consider a drop of ink of density $\rho$ moving through a lake of the same density $\rho$.
(Here, for simplicity, we assume that the fluid is incompressible so that $\rho$ is constant;
the effect of compressibility is discussed in Sec.~\ref{SECcompr}.)  
Roughly speaking, (\ref{eqp4}) implies that the ink pressure $p_{\rm ink}$ is larger than the lake 
pressure $p_{\rm lake}$ by a factor $\tilde{\gamma}^2(z,v)/\tilde{\gamma}^2(z,0)$. To make this statement 
more precise, we need to explicitly integrate the differential equation (\ref{eqp4}). While it can be 
integrated exactly, it is more illuminating to integrate it in the approximation $\rho\gg p$, in which case
(\ref{eqp4}) gives
\begin{equation}\label{dp1}
 p(z,v)-p(z_0,v)=\rho g_{\rm acc} \int_{z}^{z_0} dz\, \tilde{\gamma}^2(z,v) ,
\end{equation}
where $z_0$ is the top surface,
while $\tilde{\gamma}(z,v)$ is defined by (\ref{tgamma}) and (\ref{metric})-(\ref{potential}).
From this expression we can see that $\rho\gg p$ is a good approximation
if $g_{\rm acc} \int_{z}^{z_0} dz\, \tilde{\gamma}^2 \ll 1$. 
We assume that $v$ is not ultra-relativistic, so $\tilde{\gamma}\sim 1$ and the condition simplifies to
$g_{\rm acc} \Delta z \ll 1$. Restoring the units with the speed of light $c\neq 1$, this is really
$g_{\rm acc} \Delta z \ll c^2$. Since $c\approx 3\cdot 10^8 \,{\rm m}/{\rm s}$ and 
$g_{\rm acc} \approx 10 \, {\rm m}/{\rm s}^2$, we see that the condition $g_{\rm acc} \Delta z \ll c^2$
is indeed well satisfied for any reasonable fluid depth $\Delta z$.
Hence the pressure difference between the bottom and top of the moving ink is
\begin{equation}\label{dp2}
 \Delta p_{\rm ink}=\rho g_{\rm acc} \int_{z}^{z_0} dz\, \tilde{\gamma}^2(z,v), 
\end{equation}
while the pressure difference for the non-moving lake is 
\begin{equation}\label{dp3}
 \Delta p_{\rm lake}=\rho g_{\rm acc} \int_{z}^{z_0} dz\, \tilde{\gamma}^2(z,0) .
\end{equation}
Since $\tilde{\gamma}^2(z,v)$ increases with $|v|$, it follows that $\Delta p_{\rm ink}>\Delta p_{\rm lake}$.

\begin{figure}[t]
%\centering
\includegraphics[width=4cm]{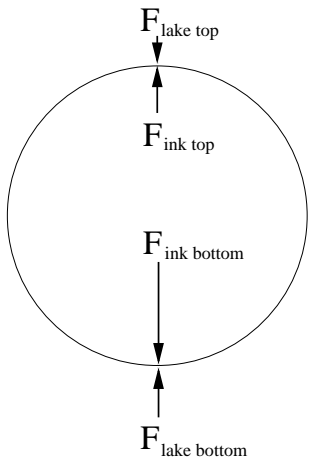}
\caption{\label{fig1}
The forces acting on the top boundary and bottom boundary of the spherical ink drop surrounded by the lake. 
The tips of arrows indicate the application points of the forces.
Since there is a force acting on each point of the boundary, the indicated forces on two points 
(the top point and the bottom point) are in fact infinitesimal, while the lengths of arrows 
indicate the relative magnitudes of the infinitesimal forces.  
Such a visualization of forces at the top and bottom of the spherical drop is sufficient to conclude
that the net force on the entire drop points downwards. 
}
\end{figure}

Now consider a surface at the boundary between the two fluids. At any point on the surface, 
the fluid with energy-momentum $T^{\mu\nu}$ acts
%on the surface 
with the local infinitesimal force \cite{hartle}
\begin{equation}\label{dF}
d F^i=T^{ij} d a_j ,
\end{equation}
where $d a_j$ is the outward oriented infinitesimal area element. 
We are interested in a horizontal surface, for which the only non-vanishing component of $d a_j$ is
$d a_3$, so (\ref{dF}) reduces to $d F^3=T^{33}d a_3$. Since $u^3=0$, (\ref{enmom}) gives $T^{33}=p$, 
so we have
\begin{equation}\label{dF2}
d F^3= p\, d a_3 ,
\end{equation}
showing that the local vertical force per unit area depends only on $p$. 
But on any surface we have in fact two forces, one coming from fluid on one side of the surface 
and another coming from fluid on the other side of the surface.
Hence the total force acting on the boundary surface is
\begin{equation}
 d F^3=d F^3_{\rm ink}+ d F^3_{\rm lake}=p_{\rm ink} d a_3^{\rm ink}+p_{\rm lake} d a_3^{\rm lake} .
\end{equation}
But $d a_3^{\rm lake}=-d a_3^{\rm ink}$, so the total force on the boundary surface 
is given by the difference of the two pressures
\begin{equation}
 d F^3=(p_{\rm ink}-p_{\rm lake}) d a_3^{\rm ink} .
\end{equation}
Since $p_{\rm ink}>p_{\rm lake}$, it follows that the force has the same direction as $d a_3^{\rm ink}$,
i.e. in the outward direction from the ink drop. Hence the force on the bottom boundary of the drop acts 
downward, while the force on the top boundary of the drop acts upward.
But the force on the bottom is stronger than the force on the top because the pressures are larger on the bottom,
as seen in (\ref{dp1})-(\ref{dp3}). 
More explicitly, after (\ref{dp3}) we have shown that
$\Delta p_{\rm ink}>\Delta p_{\rm lake}$, i.e.
\begin{equation}
 p_{\rm ink,bottom}-p_{\rm ink,top}>p_{\rm lake,bottom}-p_{\rm lake,top} ,
\end{equation}
which can be rewritten as
\begin{equation}
 (p_{\rm ink}-p_{\rm lake})_{\rm bottom} > (p_{\rm ink} -p_{\rm lake})_{\rm top} ,
\end{equation}
which shows that the downward pressure on the bottom is larger than the upward pressure on the top.
Hence the net force on the drop has the same direction as 
the force on the bottom boundary, i.e. the overall force on the drop is downward. 
For a visualization of all these forces see also Fig.~\ref{fig1}.
This implies that the drop {\em sinks}.

Some readers may not be completely convinced by the analysis of forces above
because it may differ from the way the forces are analyzed in more elementary 
texts on non-relativistic mechanics. Why isn't the mass of the drop 
involved in the analysis? Why is the gravitational force on the drop not 
considered explicitly? 
%Since the forces at the bottom and top boundary act in 
%opposite directions, why isn't the drop just stretched, rather than moved in the down direction? 
To address those issues in the simplest possible manner, let us consider 
a non-relativistic cubical drop of constant density $\rho$, 
surrounded by vacuum (not by the lake).
The cube has the height $\Delta z$ and the horizontal base with the area $A$.
The pressure difference is $\Delta p =\rho g_{\rm acc}\Delta z$, so the 
force difference $\Delta F=A \Delta p = m g_{\rm acc}$
points in the downward direction, where $m=\rho A\Delta z$
is the mass of the drop. Thus we see that the formula for pressure difference 
contains a dependence on mass implicitly, 
through the dependence on $\rho$.
Moreover, we see that the force difference $\Delta F$
is the {\em same thing} as the gravitational force $F_{\rm grav}=m g_{\rm acc}$ 
on the drop in the downward direction. 
%From the everyday experience with rain 
%we know that the shape of drop is not changed during the free fall,
%which can qualitatively be understood as follows. The drop's volume cannot change 
%because $\rho$ is constant. So if the drop was stretched in the vertical direction 
%it would have to shrink in the horizontal direction, which however is prevented 
%by the pressures on the lateral sides 
%(e.g. four lateral faces of the cubical drop) that act in the outward directions.

Let us also note that (\ref{dF}) is not manifestly covariant because it contains only spatial indices.
A manifestly covariant version of (\ref{dF}) is presented heuristically in \cite{hartle}.
For completeness, in Appendix \ref{SECapp} we present the covariant version of (\ref{dF}) in a mathematically
precise form. 
%Readers not interested in mathematical
%subtleties may skip the Appendix entirely and just keep in mind 
%that a manifestly covariant version of (\ref{dF}) exists. 
This relativistic covariance implies that 
all observers agree that the drop sinks.  

\section{The role of compressibility}  
\label{SECcompr}

So far we discussed rigid submarines and incompressible fluids, but those are idealizations.
In fact, a strictly rigid body and a strictly incompressible fluid imply that any local disturbance propagates 
with infinite speed, which is inconsistent with relativity. Hence a proper relativistic treatment of a fluid 
needs to take compressibility into account. In this section we study the effect of compressibility 
in a slightly compressible fluid like water and show that it does not significantly affect the results,
provided that the size of the fluid drop is small enough.

In an adiabatic process, the compressibility of a fluid is related to the speed of sound $c_s$ 
through the formula \cite{LL6}   
\begin{equation}
\frac{\partial p}{\partial \rho}=c_s^2 .
\end{equation}
In the approximation of a constant $c_s$, this implies the equation of state
\begin{equation}\label{rhop}
 \rho(p)=c_s^{-2}p+\rho_0 ,
\end{equation}
where $\rho_0$ is a constant not depending on $p$. 
But Eq.~(\ref{eqp4}) is valid for any equation of state $\rho(p)$, 
even when the approximation $\rho\gg p$ is not valid,
so we see that 
compressibility does not affect the conclusion that $p$ gets enhanced by the factor 
$\tilde{\gamma}^2(z,v)$ due to the fluid motion, owing to which the moving drop sinks.

The compressibility, however, can affect how much the pressure is enhanced.  
The insertion of 
(\ref{rhop}) into (\ref{eqp4}) gives
\begin{equation}\label{eqp4com}
 \partial_z p=-(p+c_s^{-2}p+\rho_0)g_{\rm acc} \tilde{\gamma}^2 .
\end{equation}
It can be integrated exactly, most easily by writing it as 
\begin{equation}
 \frac{dp}{(1+c_s^{-2})p+\rho_0}=-dz\, g_{\rm acc} \tilde{\gamma}^2(z,v) ,
\end{equation}
and integrating the left-hand side over $p$ and the right-hand side over $z$.
We now assume that the speed of sound is much smaller than the speed of light, 
i.e. $c_s\ll 1$. We also assume that the fluid is only slightly compressed,
which, due to (\ref{rhop}), implies $c_s^{-2}p\ll\rho_0$. Thus we have a hierarchy
\begin{equation}
 p\ll c_s^{-2}p \ll \rho_0 ,
\end{equation}
which shows that (\ref{eqp4com}) can be approximated with 
\begin{equation}\label{eqp4com2}
 \partial_z p=-\rho_0g_{\rm acc} \tilde{\gamma}^2 .
\end{equation}
Hence we conclude that we can again use the approximation (\ref{dp1}), with $\rho\to\rho_0$.

The conclusion above rests on the assumption $c_s^{-2}p\ll  \rho_0$. Now let us see 
under what condition is this assumption satisfied.
%by a quantitative analysis similar to that after (\ref{dp1}).
Since this assumption implies the approximation 
$\Delta p = \rho_0 g_{\rm acc} \int_{z}^{z_0} dz\, \tilde{\gamma}^2$, taking $\tilde{\gamma}\sim 1$ as before
we see that $c_s^{-2}p\ll \rho_0$ is satisfied if $c_s^{-2}\rho_0 g_{\rm acc}\Delta z \ll \rho_0$,
or equivalently, if
\begin{equation}\label{deltaz}
 \Delta z \ll \frac{c_s^{2}}{g_{\rm acc}} .
\end{equation}
For water at the temperature of $20^{\circ}$C, the speed of sound is $c_s=1.48\cdot 10^3\, {\rm m}/{\rm s}$. 
Inserting this and $g_{\rm acc} = 9.81 \, {\rm m}/{\rm s}^2$ into (\ref{deltaz}), we 
finally obtain that the assumption is satisfied if $\Delta z \ll 2\cdot 10^5 \, {\rm m}=200  \, {\rm km}$.
Those are depths at which water is not compressed much under its own weight. 

\section{The full weak field limit}
\label{SECfull}

So far we analyzed everything in the quasi-Newtonian limit (\ref{metric}) in which only $g_{00}$ is nontrivial.
Now we want to improve our analysis by considering the full weak field limit in which both $g_{00}$ and $g_{ij}$ 
are nontrivial. In the full weak field limit, the metric (\ref{metric}) is modified to \cite{carroll}
\begin{eqnarray}\label{metric_full}
 & g_{00}=-(1+2\phi), &
\nonumber \\
 & g_{0i}=0, \;\;\; g_{ij}=(1-2\phi)\delta_{ij}, & 
\end{eqnarray}
where $\phi$ is the same as in (\ref{potential}). 
The 4-velocity still has the form (\ref{ug}), but, owing to the modified metric,
the $\tilde{\gamma}$ factor is modified:
\begin{equation}\label{ug_full}
 u^{\mu}=\tilde{\gamma}(1,v,0,0) ,
\end{equation}
\begin{equation}\label{tgamma_full}
 \tilde{\gamma}=\frac{1}{\sqrt{|g_{00}|-g_{11}v^2}} .
\end{equation}
A straightforward computation shows that (\ref{eqp2}) and (\ref{Christ}) modify to
\begin{equation}\label{eqp2_full}
g^{33}\partial_{3}p+ (p+\rho)
\left[\Gamma^{3}_{00}u^{0}u^{0}+2\Gamma^{3}_{01}u^{0}u^{1}+\Gamma^{3}_{11}u^{1}u^{1} \right]=0 , 
\end{equation}
\begin{eqnarray}\label{Christ_full}
 & \Gamma^{3}_{00}=-\displaystyle\frac{g^{33}}{2}\partial_3g_{00}, &
\nonumber \\
 & \Gamma^{3}_{01}=\Gamma^{3}_{10}=0, \;\;\; \Gamma^{3}_{11}=-\displaystyle\frac{g^{33}}{2}\partial_3g_{11} . &
\end{eqnarray}
%where we have used $\Gamma^{3}_{01}=\Gamma^{3}_{10}$. 
Thus we see that all non-vanishing terms in (\ref{eqp2_full}) are proportional to $g^{33}$, so 
the dependence on $g^{33}$ is eliminated by multiplying 
(\ref{eqp2_full}) by $1/g^{33}$. We also see that (\ref{derg00}) generalizes to 
\begin{equation}\label{derg00_full}
 \frac{\partial_3g_{00}}{2}=-g_{\rm acc} , \;\;\;
 \frac{\partial_3g_{11}}{2}=-g_{\rm acc} , 
\end{equation}
so (\ref{eqp2_full}) simplifies to
\begin{equation}
 \partial_{3}p+(p+\rho)g_{\rm acc} [(u^0)^2+(u^1)^2] =0.
\end{equation}
Using also (\ref{ug_full}), this finally can be written as
\begin{equation}\label{final_full}
 \partial_{3}p=-(p+\rho)g_{\rm acc} \tilde{\Gamma}^2 ,
\end{equation}  
where
\begin{equation}\label{tGamma}
 \tilde{\Gamma}^2=\tilde{\gamma}^2 (1+v^2) .
\end{equation}
Eq.~(\ref{final_full}) is a modification of (\ref{eqp4}). In (\ref{tGamma}), both factors $\tilde{\gamma}^2$ and $(1+v^2)$
increase with $v$, so $\tilde{\Gamma}^2$ in (\ref{final_full}) increases with $v$, similarly to  $\tilde{\gamma}^2$ in (\ref{eqp4}).
This shows that (\ref{final_full}) leads to similar qualitative behavior as (\ref{eqp4}), i.e. that 
the full weak field limit predicts an increase of pressure with velocity, similar to the quasi-Newtonian limit.

\section{Discussion}  
\label{SECdisc}

Our results can be qualitatively 
summarized as follows. When the fluid moves with horizontal velocity $v$, its pressure $p$ gets enhanced
by the $v$-dependent factor $\tilde{\gamma}^2$ defined by (\ref{tgamma}). But $p$ is a scalar, so its value does not depend 
on the choice of coordinates. Hence the pressure is enhanced by the same factor even in the comoving frame, in which
the fluid does not have velocity $v$. It may seem
paradoxical because the fluid's velocity vanishes in the comoving frame
leading one to believe that the enhancement factor cannot be the same.
The apparent paradox is resolved by noting that the effect depends on the velocity {\em relative to Earth}, not on velocity relative 
to the observer. All observers agree that the fluid moves with respect to 
the gravitational field produced by Earth. This is why the $v$-dependent enhancement does not depend 
on the velocity of the observer. This $v$-dependent enhancement of the pressure implies that the moving drop has a 
larger pressure than the surrounding non-moving lake, due to which the drop sinks. This is the essence of the resolution 
of our softened version of the submarine paradox.

Our results are in a qualitative agreement with similar results \cite{supplee,matsas,jonsson,vieira} for a rigid submarine.
In particular, in a special-relativistic treatment in \cite{supplee}, the downward pressure of the moving submarine is
enhanced by the special-relativistic factor $\gamma^2$ defined by (\ref{gamma}), with one $\gamma$ coming from Lorentz contraction
and the other $\gamma$ coming from $v$-dependent 
``relativistic mass'', 
the consequence of which is that the submarine sinks.
Similar results for a rigid submarine are obtained also in other works \cite{matsas,jonsson,vieira}.

Since the effect depends on the velocity $v$ relative to Earth, we can also understand immediately what happens 
if instead of a lake we have a fast river moving with horizontal velocity $v$ with respect to Earth, while the ink drop does 
not move with respect to Earth. In this case we have $p_{\rm river}>p_{\rm ink}$, so the ink drop floats. 
Likewise, one can also consider a symmetric situation in which the river has velocity $v/2$ with respect to Earth
and the drop has velocity $-v/2$ with respect to Earth, in which case the drop neither sinks nor floats.

Note also that our results are not in contradiction with the non-relativistic Bernoulli's effect which states
that a moving fluid has a {\em lower} pressure. In our analysis we tacitly assume that fluid is put into 
a state of larger velocity by acting on it with an {\em external} force which gives additional energy to the fluid, 
while the Bernoulli's effect assumes that there is no such external force.

Our results offer a new perspective on the submarine paradox.
In our approach the rigid submarine is replaced by a drop of ideal fluid, 
which simplifies the analysis to a certain extent, while the nature of the paradox and final results 
are qualitatively similar.
We believe that such a perspective contributes to overall understanding of the whole issue. 
 
\section*{Author Declarations}

The authors have no conflicts to disclose.
   
\section*{Acknowledgments}

The author is grateful to N. Bili\'c, G. Duplan\v{c}i\'c, T. Juri\'c, T. Maudlin and I. Smoli\'c for
discussions and/or comments on the manuscript.
This work was supported  
by the Ministry of Science of the Republic of Croatia.

\appendix

\section{The covariant version of Eq.~(\ref{dF})}
\label{SECapp}

Here we study the manifestly covariant version of (\ref{dF}).
We use the same physical ideas as \cite{hartle}, but present them 
in a form mathematically more precise than in \cite{hartle}.
This Appendix uses a higher level of mathematical sophistication
than the rest of the paper, which makes it suitable to advanced readers.

Consider a timelike 3-dimensional hypersurface embedded in the 4-dimensional spacetime.
The embedding is defined by 4 functions $x^{\mu}(y^0,y^1,y^2)$, where $y^0,y^1,y^2$ are coordinates on the 
hypersurface. The induced metric on the hypersurface is
\begin{equation}
 q_{ab}= \frac{\partial x^{\mu}}{\partial y^{a}} \frac{\partial x^{\mu}}{\partial y^{b}} g_{\mu\nu} ,
\end{equation}
where $a,b\in\{0,1,2\}$. The volume element on the hypersurface is $d^3y\sqrt{q}$, where $d^3y=dy^0dy^1dy^2$
and $q$ is the absolute value of the determinant of $q_{ab}$. 
From the 4-dimensional point of view, this 3-dimensional volume element is represented by the 4-vector
\begin{equation}\label{Asigma}
 \Delta\Sigma_{\nu}=d^3y\sqrt{q}\, n_{\nu} ,
\end{equation}
where $n_{\nu}$ is the unit spacelike vector normal to the timelike hypersurface. 

Now consider a 2-dimensional surface with coordinates $y^1,y^2$ at a fixed time $y^0$.
Let $\delta P^{\mu}$ be the amount of 4-momentum that, during a finite time,
crosses an infinitesimal part of the surface with area proportional to $dy^1dy^2$.
Then during the infinitesimal time $dy^0$, the amount of 4-momentum that crosses this part of surface is 
$d(\delta P^{\mu})\equiv \Delta P^{\mu}$.   
In terms of the quantities defined as above, the covariant version of (\ref{dF}) is
\begin{equation}\label{dFcov}
\Delta P^{\mu}=T^{\mu\nu}\Delta\Sigma_{\nu} .
\end{equation}

To see more explicitly how is (\ref{dFcov}) related to (\ref{dF}), we choose coordinates $y^a$ on the hypersurface
such that $q_{01}=q_{02}=0$. With this choice, the determinant $q$ factorizes as $q=|q_{00}|\,q^{(2)}$, where 
$q^{(2)}$ is the determinant of the induced metric tensor on the 2-dimensional surface. Thus we can write  
(\ref{Asigma}) as
\begin{equation}\label{Asigma2}
\Delta\Sigma_{\nu}=dt_{\rm loc} \delta a_{\nu} ,
\end{equation}
where 
\begin{equation}
dt_{\rm loc}=dy^0\sqrt{|q_{00}|}, \;\;\; \delta a_{\nu}=d^2y\sqrt{q^{(2)}} \, n_{\nu} ,
\end{equation}
and $d^2y=dy^1dy^2$. Hence (\ref{dFcov}) can be written as 
\begin{equation}\label{dFcov2}
\frac{\Delta P^{\mu}}{dt_{\rm loc}}=T^{\mu\nu}\delta a_{\nu} .
\end{equation}
The left-hand side can be written as
\begin{equation}
\frac{\Delta P^{\mu}}{dt_{\rm loc}} = \frac{d(\delta P^{\mu})}{dt_{\rm loc}} 
= \delta\left( \frac{dP^{\mu}}{dt_{\rm loc}} \right) ,
\end{equation}
so (\ref{dFcov2}) becomes
\begin{equation}\label{dFcov3}
\delta\left( \frac{dP^{\mu}}{dt_{\rm loc}} \right)=T^{\mu\nu}\delta a_{\nu} .
\end{equation}

Eq.~(\ref{dFcov3}) is quite similar to (\ref{dF}), but to get (\ref{dF}) we would like to eliminate 
the contribution from $\nu=0$. It could be done by an appropriate choice of 4-dimensional spacetime coordinates, but
we want to keep the 4-dimensional spacetime coordinates arbitrary.
Therefore we choose a different method, based on tetrad formalism \cite{tetrad}. 
The tetrad $e_{\bar{\alpha}}^{\mu}(x)$ is an object with two kinds of indices, 
the index $\mu$ in the curved spacetime and the index $\bar{\alpha}$ in the flat tangential space at $x$.
It obeys $e_{\bar{\alpha}}^{\mu} e_{\bar{\beta}}^{\nu} g_{\mu\nu}=\eta_{\bar{\alpha}\bar{\beta}}$, 
where $\eta_{\bar{\alpha}\bar{\beta}}$ is the flat Minkowski metric. From any object with spacetime indices 
one can construct an equivalent object with Minkowski indices, e.g. $A_{\bar{\alpha}}= e_{\bar{\alpha}}^{\mu}A_{\mu}$ 
and $A^{\bar{\alpha}}= e^{\bar{\alpha}}_{\mu}A^{\mu}$, where 
$e^{\bar{\alpha}}_{\mu}=\eta^{\bar{\alpha}\bar{\beta}} g_{\mu\nu} e_{\bar{\beta}}^{\nu}$.
With this formalism we can write (\ref{dFcov3}) as
\begin{equation}\label{dFcov4}
\delta\left( \frac{dP^{\bar{\mu}}}{dt_{\rm loc}} \right)=T^{\bar{\mu}\bar{\nu}}\delta a_{\bar{\nu}} .
\end{equation}
The advantage of (\ref{dFcov4}) over (\ref{dFcov3}) is that now we can choose the tetrad such that 
the spacelike vector $n_{\bar{\nu}}$ has $n_{\bar{0}}=0$, so (\ref{dFcov4}) reduces to 
\begin{equation}\label{dFcov5}
\delta\left( \frac{dP^{\bar{\mu}}}{dt_{\rm loc}} \right)=T^{\bar{\mu}\bar{j}}\delta a_{\bar{j}} ,
\end{equation} 
with $\bar{j}\in\{ 1,2,3\}$. 
Choosing the tetrad in a specific way while keeping coordinates arbitrary,
rather than the other way around, makes the formalism manifestly general covariant.
Finally, choosing $\bar{\mu}=\bar{i}$ and defining the infinitesimal force as
\begin{equation}
 \delta F^{\bar{i}}=\delta\left( \frac{dP^{\bar{i}}}{dt_{\rm loc}} \right),
\end{equation}
(\ref{dFcov5}) implies
\begin{equation}\label{dFcov6}
\delta F^{\bar{i}}=T^{\bar{i}\bar{j}}\delta a_{\bar{j}} ,
\end{equation}
which is nothing but (\ref{dF}) written in a more precise form.


\begin{thebibliography}{99}
\bibitem{footnote}
The full weak field limit depends on the Einstein equation
which specifies the relation between spacetime curvature and matter energy-momentum tensor.
The quasi-Newtonian limit, on the other hand, is the simplest modification of 
flat Minkowski metric that yields Newtonian gravity for small velocities;
it does not depend on the Einstein equation, but
only on the general kinematic idea that gravity is described by spacetime metric.
\end{thebibliography}
\end{document}